\documentclass[aps,prd,preprint,superscriptaddress,tightenlines,%
nofootinbib]{revtex4}
\newcommand{\PRE}[1]{{#1}}   

\usepackage{bm}
\usepackage{epsfig}
\usepackage{latexsym}

\newcommand{\postscript}[2]{\setlength{\epsfxsize}{#2\hsize}
   \centerline{\epsfbox{#1}}}
\newcommand{\comment}[1]{}

\def\comment#1{{}}

\def\slashchar#1{\setbox0=\hbox{$#1$}           
   \dimen0=\wd0                                 
   \setbox1=\hbox{/} \dimen1=\wd1               
   \ifdim\dimen0>\dimen1                        
      \rlap{\hbox to \dimen0{\hfil/\hfil}}      
      #1                                        
   \else                                        
      \rlap{\hbox to \dimen1{\hfil$#1$\hfil}}   
      /                                         
   \fi}
\newif\ifnref

\nreffalse

\input epsf
\def\figin{\epsfcheck\figin}\def\figins{\epsfcheck\figins}
\def\epsfcheck{\ifx\epsfbox\UnDeFiNeD
\message{(NO epsf.tex, FIGURES WILL BE IGNORED)}
\gdef\figin##1{\vskip2in}\gdef\figins##1{\hskip.5in}
\else\message{(FIGURES WILL BE INCLUDED)}%
\gdef\figin##1{##1}\gdef\figins##1{##1}\fi}
\def\DefWarn#1{}
\def\figinsert{\goodbreak\midinsert}
\def\ifig#1#2#3{\DefWarn#1\xdef#1{fig.~\the\figno}
\writedef{#1\leftbracket fig.\noexpand~\the\figno}%
\figinsert\figin{\centerline{#3}}\medskip\centerline{\vbox{\baselineskip12pt
\advance\hsize by -1truein\noindent\footnotefont{\bf Fig.~\the\figno } #2}}
\bigskip\endinsert\global\advance\figno by1}

\begin{document}

\preprint{
\hfil
\begin{minipage}[t]{3in}
\begin{flushright}
\vspace*{.4in}
MPP--2008--86\\
LMU-ASC 42/08\\
NUB-3262-Th-08\\
\end{flushright}
\end{minipage}
}

\vspace{1cm}

\title{Dijet signals for low mass strings at the LHC
\PRE{\vspace*{0.3in}} }

\author{Luis A. Anchordoqui}
\affiliation{Department of Physics,\\
University of Wisconsin-Milwaukee,
 Milwaukee, WI 53201, USA
\PRE{\vspace*{.1in}}
}

\author{Haim Goldberg}
\affiliation{Department of Physics,\\
Northeastern University, Boston, MA 02115, USA
\PRE{\vspace*{.1in}}
}

\author{Dieter L\"ust}
\affiliation{Max--Planck--Institut f\"ur Physik\\
 Werner--Heisenberg--Institut,
80805 M\"unchen, Germany
\PRE{\vspace*{.1in}}
}

\affiliation{Arnold Sommerfeld Center for Theoretical Physics\\
Ludwig-Maximilians-Universit\"at M\"unchen,
80333 M\"unchen, Germany
\PRE{\vspace{.1in}}
}

\author{Satoshi \nolinebreak Nawata}
\affiliation{Department of Physics,\\
University of Wisconsin-Milwaukee,
 Milwaukee, WI 53201, USA
\PRE{\vspace*{.1in}}
}

\author{Stephan Stieberger}
\affiliation{Max--Planck--Institut f\"ur Physik\\
 Werner--Heisenberg--Institut,
80805 M\"unchen, Germany
\PRE{\vspace*{.1in}}
}

\author{Tomasz R. Taylor}
\affiliation{Department of Physics,\\
  Northeastern University, Boston, MA 02115, USA \PRE{\vspace*{.1in}}
}

\date{August 2008}
\PRE{\vspace*{.5in}}
\begin{abstract}\vskip 3mm
  \noindent
  We consider extensions of the standard model based on open strings
  ending on D-branes, with gauge bosons due to strings attached to
  stacks of D-branes and chiral matter due to strings stretching
  between intersecting D-branes.  Assuming that the fundamental string
  mass scale is in the TeV range and the theory is weakly coupled, we
  discuss possible signals of string physics at the Large Hadron
  Collider (LHC).  In such D-brane constructions, the dominant
  contributions to full-fledged string amplitudes for all the common
  QCD parton subprocesses leading to dijets are completely independent
  of the details of compactification, and can be evaluated in a
  parameter-free manner. We make use of these amplitudes evaluated
  near the first resonant pole to determine the discovery potential of
  LHC for the first Regge excitations of the quark and
  gluon. Remarkably, the reach of LHC after a few years of running can
  be as high as 6.8~TeV. Even after the first 100~pb$^{-1}$ of
  integrated luminosity, string scales as high as 4.0~TeV can be
  discovered. For string scales as high as 5.0~TeV, observations of
  resonant structures in $pp\rightarrow {\rm direct}\ \gamma~ +$ jet
  can provide interesting corroboration for string physics at the
  TeV-scale.

\end{abstract}

\maketitle

Experiments at the Large Hadron Collider (LHC) will search for the
Higgs boson -- the last missing piece of the standard model.  At the
same time, searches will be conducted for signals of new physics
beyond it. Although many extensions of the standard model have been
proposed over the last thirty years, none of them has been favored so
far by the experimental data. There are differences, however, between
various models with respect to their testability in the energy range
accessible at the LHC. It has recently become
clear~\cite{Anchordoqui:2007da,Anchordoqui:2008ac,Anchordoqui:2008hi,stringhunter}
that superstring theory offers one of the most robust scenarios beyond
the standard model, provided that its fundamental mass scale is
sufficiently ``low'', {\it i.e}.\ of order TeV.  The purpose of this
Letter is to discuss dijet signals of low mass string theory at the
LHC and to determine the discovery reach based on the measurements of
the corresponding cross sections.

The mass scale $M_s$ of fundamental strings can be as low as few TeV
provided that spacetime extends into large extra dimensions, allowing
a novel solution for the hierarchy problem~\cite{add,ant}. This mass
determines the center of mass energy threshold $\sqrt{s}\ge M_s$ for
the production of Regge resonances in parton collisions, thus for the
onset of string effects at the LHC~\cite{stringhunter}. (Mandelstam
variables $s,$ $t,$ $u$ used in this Letter refer to parton
subprocesses.) We consider the extensions of the standard model based
on open strings ending on D-branes, with gauge bosons due to strings
attached to stacks of D-branes and chiral matter due to strings
stretching between intersecting D-branes~\cite{inter}. Only one
assumption is necessary in order to set up a solid framework: the
string coupling must be small in order to rely on perturbation theory
in the computations of scattering amplitudes. In this case, black hole
production and other strong gravity effects occur at energies above
the string scale; therefore at least a few lowest Regge recurrences
are available for examination, free from interference with some
complex quantum gravitational phenomena.  Starting from a small string
coupling, the values of standard model coupling constants are
determined by D-brane configurations and the properties of extra
dimensions, hence that part of superstring theory requires intricate
model-building; however, as argued in
Refs.~\cite{Anchordoqui:2007da,Anchordoqui:2008ac,Anchordoqui:2008hi,stringhunter},
some basic properties of Regge resonances like their production rates
and decay widths are completely model-independent.  The resonant
character of parton cross sections should be easy to detect at the LHC
if the string mass scale is not too high.  Direct photon production
channels discussed in~\cite{Anchordoqui:2007da,Anchordoqui:2008ac} are
very interesting because the signal is due to processes that are
absent in the standard model at the leading (tree level) order of
perturbation theory. On the other hand, dijet production is a standard
tree-level QCD process. However, as argued in this Letter, the
presence of string resonances may lead to a dramatic enhancement of
the production rates and allow access to higher string mass
scales. In addition, dijet calculations do not depend on the unknown  mixing 
parameter characterizing the embedding of hypercharge in the extended D-brane 
gauge group. Moreover, as discussed later in this work, future measurements
of dijet angular distributions can provide a potent method for
distinguishing between various compactification scenarios.

The physical processes underlying dijet production at the LHC are the
collisions of two partons, producing two final partons that fragment
into hadronic jets. The corresponding $2\to 2$ scattering amplitudes,
computed at the leading order in string perturbation theory, are
collected in Ref.~\cite{stringhunter}. The amplitudes involving four
gluons as well as those with two gluons plus two quarks do not depend
on the compactification details and are completely model-independent.
All string effects are encapsulated in these amplitudes in one
``form factor'' function of Mandelstam variables $s,~t,~u$ (constrained
by $s+t+u=0$):
\begin{equation}
V(  s,   t,   u)= \frac{s\,u}{tM_s^2}B(-s/M_s^2,-u/M_s^2)={\Gamma(1-   s/M_s^2)\ \Gamma(1-   u/M_s^2)\over
    \Gamma(1+   t/M_s^2)}.\label{formf}
\end{equation}
The physical content of the form factor becomes clear after using the
well-known expansion in terms of $s$-channel resonances
\cite{gabriele}:
\begin{equation}
B(-s/M_s^2,-u/M_s^2)=-\sum_{n=0}^{\infty}\frac{M_s^{2-2n}}{n!}\frac{1}{s-nM_s^2}
\Bigg[\prod_{J=1}^n(u+M^2_sJ)\Bigg],\label{bexp}
\end{equation}
which exhibits $s$-channel poles associated to the propagation of
virtual Regge excitations with masses $\sqrt{n}M_s$. Thus near the
$n$th level pole $(s\to nM^2_s)$:
\begin{equation}\qquad
V(  s,   t,   u)\approx \frac{1}{s-nM^2_s}\times\frac{M_s^{2-2n}}{(n-1)!}\prod_{J=0}^{n-1}(u+M^2_sJ)\ .\label{nthpole}
\end{equation}
In specific amplitudes, the residues combine with the remaining
kinematic factors, reflecting the spin content of particles exchanged
in the $s$-channel, ranging from $J=0$ to $J=n+1$.

The amplitudes for the four-fermion processes like quark-antiquark
scattering are more complicated because the respective form factors
describe not only the exchanges of Regge states but also of heavy
Kaluza-Klein and winding states with a model-dependent spectrum
determined by the geometry of extra dimensions. Fortunately, they are
suppressed, for two reasons. First, the QCD $SU(3)$ color group
factors favor gluons over quarks in the initial state. Second, the
parton luminosities in proton-proton collisions at the LHC, at the
parton center of mass energies above~1 TeV, are significantly lower
for quark-antiquark subprocesses than for gluon-gluon and
gluon-quark~\cite{Anchordoqui:2008ac}. The collisions of valence
quarks occur at higher luminosity; however, there are no Regge
recurrences appearing in the $s$-channel of quark-quark
scattering~\cite{stringhunter}.

Before proceeding, we pause to present our notation. The first Regge
excitations of the gluon $(g)$ and quarks $(q)$ will be denoted by
$g^*,\ q^*$, respectively. In the D-brane models under consideration,
the ordinary $SU(3)$ color gauge symmetry is extended to $U(3)$, so
that the open strings terminating on the stack of ``color'' branes
contain an additional $U(1)$ gauge boson $C$ and its excitations to
accompany the gluon and its excitations. The first excitation of the
$C$ will be denoted by $C^*$.

In the following we isolate the contribution to the partonic cross
section from the first resonant state. Note that far below the string
threshold, at partonic center of mass energies $\sqrt{s}\ll M_s$, the
form factor $V(s,t,u)\approx
1-\frac{\pi^2}{6}{su}/M^4_s$~\cite{stringhunter} and therefore the
contributions of Regge excitations are strongly suppressed. The
$s$-channel pole terms of the average square amplitudes contributing
to dijet production at the LHC can be obtained from the general
formulae given in Ref.~\cite{stringhunter}, using
Eq.(\ref{nthpole}). However, for phenomenological purposes, the poles
need to be softened to a Breit-Wigner form by obtaining and utilizing
the correct {\em total} widths of the
resonances~\cite{Anchordoqui:2008hi}. After this is done, the
contributions of the various channels are as follows:
\begin{eqnarray}
|{\cal M} (gg \to gg)| ^2 & = & \frac{19}{12} \
\frac{g^4}{M_s^4} \left\{ W_{g^*}^{gg \to gg} \, \left[\frac{M_s^8}{(  s-M_s^2)^2
+ (\Gamma_{g^*}^{J=0}\ M_s)^2} \right. \right.
\left. +\frac{  t^4+   u^4}{(  s-M_s^2)^2 + (\Gamma_{g^*}^{J=2}\ M_s)^2}\right] \nonumber \\
   & + &
W_{C^*}^{gg \to gg} \, \left. \left[\frac{M_s^8}{(  s-M_s^2)^2 + (\Gamma_{C^*}^{J=0}\ M_s)^2} \right.
\left. +\frac{  t^4+  u^4}{(  s-M_s^2)^2 + (\Gamma_{C^*}^{J=2}\ M_s)^2}\right] \right\},
\label{gggg2}
\end{eqnarray}
\begin{eqnarray}
|{\cal M} (gg \to q \bar q)|^2 & = & \frac{7}{24} \frac{g^4}{M_s^4}\ N_f\
\left [W_{g^*}^{gg \to q \bar q}\, \frac{  u   t(   u^2+   t^2)}{(  s-M_s^2)^2 + (\Gamma_{g^*}^{J=2}\ M_s)^2} \right. \nonumber \\
 & + &  \left. W_{C^*}^{gg \to q \bar q}\, \frac{  u   t (   u^2+   t^2)}{(  s-M_s^2)^2 +
(\Gamma_{C^*}^{J=2}\ M_s)^2} \right]
\end{eqnarray}
\begin{eqnarray}
|{\cal M} (q \bar q \to gg)|^2  & = &  \frac{56}{27} \frac{g^4}{M_s^4}\
\left[ W_{g^*}^{q\bar q \to gg} \,  \frac{  u   t(   u^2+   t^2)}{(  s-M_s^2)^2 + (\Gamma_{g^*}^{J=2}\ M_s)^2} \right. \nonumber \\
 & + & \left.  W_{C^*}^{q\bar q \to gg} \, \frac{  u   t(   u^2+   t^2)}{(  s-M_s^2)^2 + (\Gamma_{C^*}^{J=2}\ M_s)^2} \right] \,\,,
\end{eqnarray}
\begin{equation}
|{\cal M}(qg \to qg)|^2  =  - \frac{4}{9} \frac{g^4}{M_s^2}\
\left[ \frac{M_s^4   u}{(  s-M_s^2)^2 + (\Gamma_{q^*}^{J=1/2}\ M_s)^2} + \frac{u^3}{(s-M_s^2)^2 + (\Gamma_{q^*}^{J=3/2}\ M_s)^2}\right],
\label{qgqg2}
\end{equation}
where $g$ is the QCD coupling constant $(\alpha_{\rm QCD}=\frac{g^2}{4\pi}\approx 0.1)$
 and $\Gamma_{g^*}^{J=0} = 75\, (M_s/{\rm TeV})~{\rm GeV}$,
$\Gamma_{C^*}^{J=0} = 150 \, (M_s/{\rm TeV})~{\rm GeV}$,
$\Gamma_{g^*}^{J=2} = 45 \, (M_s/{\rm TeV})~{\rm GeV}$,
$\Gamma_{C^*}^{J=2} = 75 \, (M_s/{\rm TeV})~{\rm GeV}$,
$\Gamma_{q^*}^{J=1/2} = \Gamma_{q^*}^{J=3/2} = 37\, (M_s/{\rm
  TeV})~{\rm GeV}$ are the total decay widths for intermediate states
$g^*$, $C^*$, and $q^*$ (with angular momentum
$J$)~\cite{Anchordoqui:2008hi}. The associated weights of these
intermediate states are given in terms of the probabilities for the
various entrance and exit channels
\begin{equation}
W_{g^*}^{gg \to gg} = \frac{8(\Gamma_{g^* \to gg})^2}{8(\Gamma_{g^* \to gg})^2 +
(\Gamma_{C^* \to gg})^2} = 0.44 \,,
\label{w1}
\end{equation}
\begin{equation}
W_{C^*}^{gg \to gg} = \frac{(\Gamma_{C^*
  \to gg})^2}{8(\Gamma_{g^* \to gg})^2 + (\Gamma_{C^* \to gg})^2} =
0.56 \, ,
\label{w2}
\end{equation}
\begin{equation}
W_{g^*}^{gg \to q \bar q}  = W_{g^*}^{q \bar q \to gg} =
\frac{8\,\Gamma_{g^* \to gg} \,
\Gamma_{g^* \to q \bar q}} {8\,\Gamma_{g^* \to gg} \,
\Gamma_{g^* \to q \bar q} + \Gamma_{C^* \to gg} \,
\Gamma_{C^* \to q \bar q}} = 0.71 \, ,
\label{w3}
\end{equation}
\begin{equation}
W_{C^*}^{gg \to q \bar q} = W_{C^*}^{q \bar q \to gg}  =
\frac{\Gamma_{C^* \to gg} \,
\Gamma_{C^* \to q \bar q}}{8\,\Gamma_{g^* \to gg} \,
\Gamma_{g^* \to q \bar q} + \Gamma_{C^* \to gg} \,
\Gamma_{C^* \to q \bar q}} = 0.29 \, .
\label{w4}
\end{equation}
Superscripts $J=2$ are understood to be inserted on all the $\Gamma$'s in
Eqs.(\ref{w1}), (\ref{w2}), (\ref{w3}), (\ref{w4}).
Equation~(\ref{gggg2}) reflects the fact that weights for $J=0$ and
$J=2$ are the same~\cite{Anchordoqui:2008hi}. In what follows we set
the number of flavors $N_f =6$.

\begin{figure}[tbp]
\begin{minipage}[t]{0.49\textwidth}
\postscript{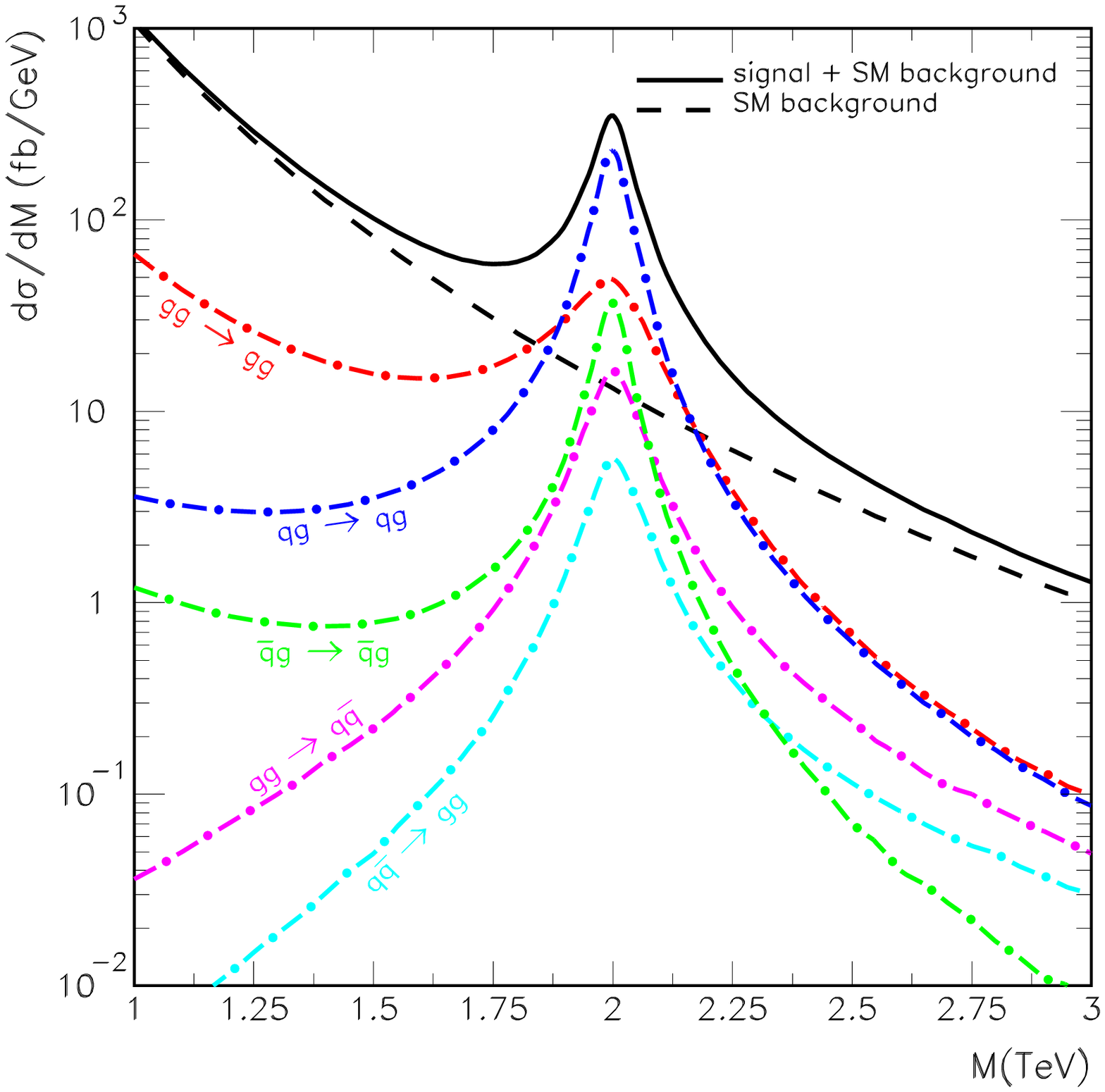}{0.99}
\end{minipage}
\hfill
\begin{minipage}[t]{0.49\textwidth}
\postscript{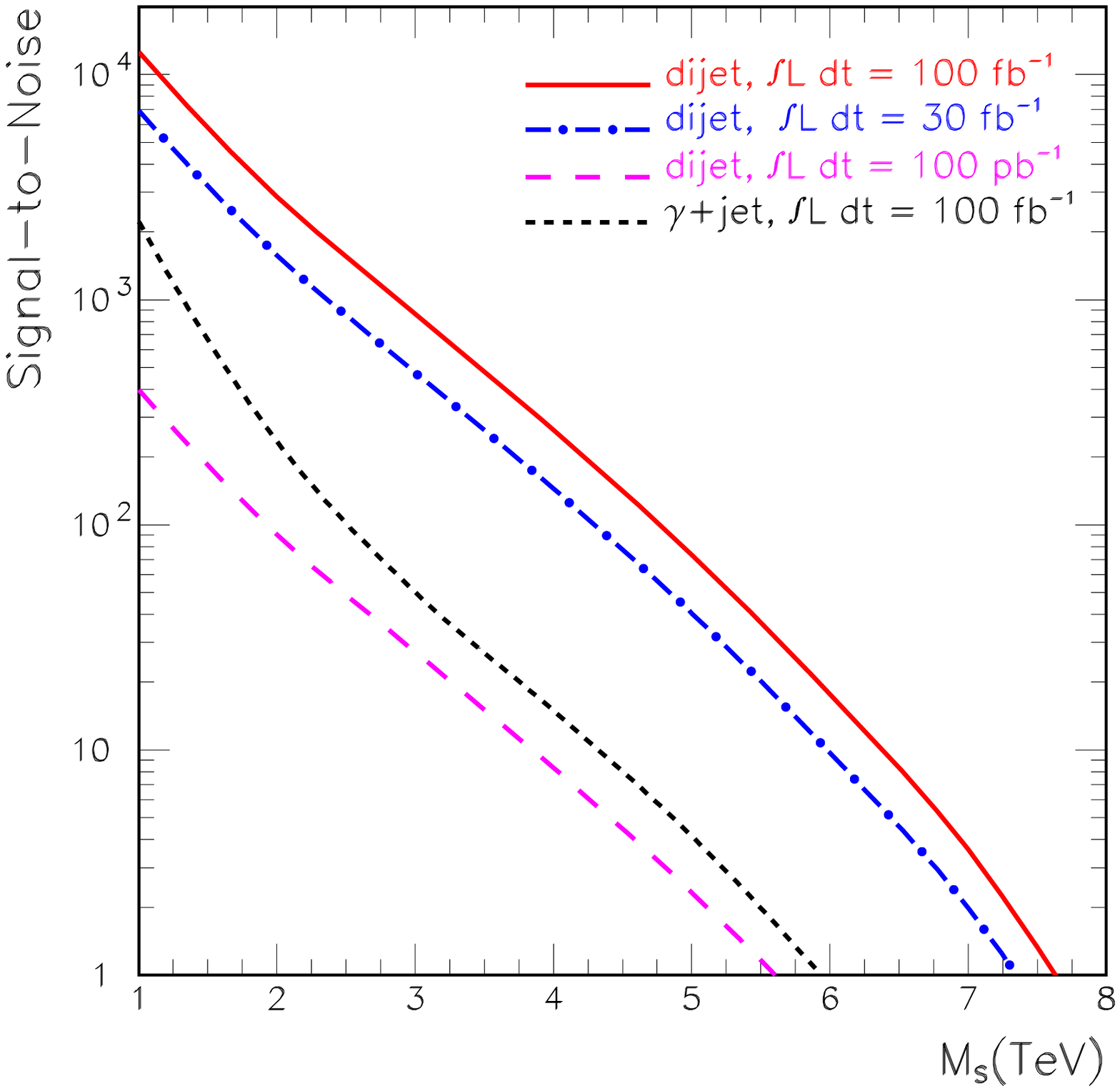}{0.99}
\end{minipage}
\caption{{\em Left panel:} $d\sigma/dM$ (units of fb/GeV) {\em vs.}
  $M$ (TeV) is plotted for the case of SM QCD background (dashed line)
  and (first resonance) string signal + background (solid line). The
  dot-dashed lines indicate the different contributions to the string
  signal ($gg \to gg$, $gg \to q \bar q$, $qg \to qg$, and $q \bar q
  \to gg$).  {\em Right panel:} $pp \to {\rm dijet}$ signal-to-noise
  ratio for three integrated luminosities. For comparison, we also
  show the signal-to-noise of $pp \to \gamma + {\rm jet}$, for
  $\kappa^2 \simeq 0.02$, see Ref.~\cite{Anchordoqui:2007da}.}
\label{fig:bump}
\end{figure}

The resonance would be visible in data binned according to the
invariant mass $M$ of the dijet, after setting cuts on the different
jet rapidities, $|y_1|, \, |y_2| \le 1$~\cite{CMS} and transverse
momenta $p_{\rm T}^{1,2}>50$ GeV.  In Fig.~\ref{fig:bump} we show a
representative plot of the invariant mass spectrum, for $M_s =2$~TeV,
detailing the contribution of each subprocess.  The QCD background has
been calculated at the partonic level from the same processes as
designated for the signal, with the addition of $qq\to qq$ and $q \bar
q \to q \bar q$.  Our calculation, making use of the CTEQ6D parton
distribution functions~\cite{Pumplin:2002vw} agrees with that
presented in~\cite{CMS}.

We now estimate (at the parton level) the LHC discovery
reach. Standard bump-hunting methods, such as obtaining cumulative
cross sections, $\sigma (M_0) = \int_{M_0}^\infty \frac{d\sigma}{dM}
\, \, dM$, from the data and searching for regions with significant
deviations from the QCD background, may reveal an interval of $M$
suspected of containing a bump.  With the establishment of such a
region, one may calculate a signal-to-noise ratio, with the signal
rate estimated in the invariant mass window $[M_s - 2 \Gamma, \, M_s +
2 \Gamma]$. The noise is defined as the square root of the number of
background events in the same dijet mass interval for the same
integrated luminosity.

The top two and bottom curves in Fig.~\ref{fig:bump}  
show the behavior
of the signal-to-noise (S/N) ratio as a function of the string scale
for three integrated luminosities (100~fb$^{-1},$ 30~fb$^{-1}$ and
100~pb$^{-1}$) at the LHC. It is remarkable that within 1-2 years of
data collection, {\it string scales as large as 6.8 TeV are open to
  discovery at the $\geq 5\sigma$ level.} For 30~fb$^{-1},$ the
presence of a resonant state with mass as large as 5.7 TeV can provide a
signal of convincing significance $(S/N = 592/36 > 13)$. The bottom curve,
corresponding to data collected in a very early run of 100~pb$^{-1},$
shows that a resonant mass as large as 4.0~TeV can be observed with
$10\sigma$ significance! Once more, we stress that these results
contain no unknown parameters. They depend only on the D-brane
construct for the standard model, and {\it are independent of
  compactification details.}

For comparison with our previous analysis, we also show in 
Fig.~\ref{fig:bump} a fourth curve, for the process $pp\rightarrow
\gamma +$ jet. (In what follows, $\gamma$ refers to an isolated 
gamma ray.) In Ref.~\cite{Anchordoqui:2008ac} a cut $(p_T^\gamma
> 300$~GeV) was selected for discovery of new physics. As far as the
signal is concerned, this cut is largely equivalent to selecting on
$\gamma$-jet invariant masses in the 2-5~TeV range, with cuts on
photon and jet rapidities $|y_1|, \, |y_2| <
2.4$~\cite{Ball:2007zza}. However, for $M_s > 2$~TeV the background
is greatly reduced with the dijet mass method used here, resulting
in an extension of the discovery reach, up to about
5~TeV~\cite{pizero}. The signal used to obtain the results displayed
in Fig.~\ref{fig:bump} includes the parton subprocesses $gg \to g
\gamma$ (which does not exist at tree level in QCD, and which was
the only subprocess evaluated
in~\cite{Anchordoqui:2007da,Anchordoqui:2008ac}), $qg \to q \gamma$,
$\bar q g \to \bar q \gamma$, and $q \bar q \to g \gamma$. All
except the first have been calculated in QCD and constitute the
standard model background. The projection of the photon onto the $C$
gauge boson was also effected in the last-cited references. Although
the discovery reach is not as high as that for dijets, the
measurement of $pp\rightarrow\gamma~ +$ jet can potentially provide
an interesting corroboration for the stringy origin for new physics
manifest as a
resonant structure in LHC data.

\begin{figure}[tbp]
\begin{minipage}[t]{0.49\textwidth}
\postscript{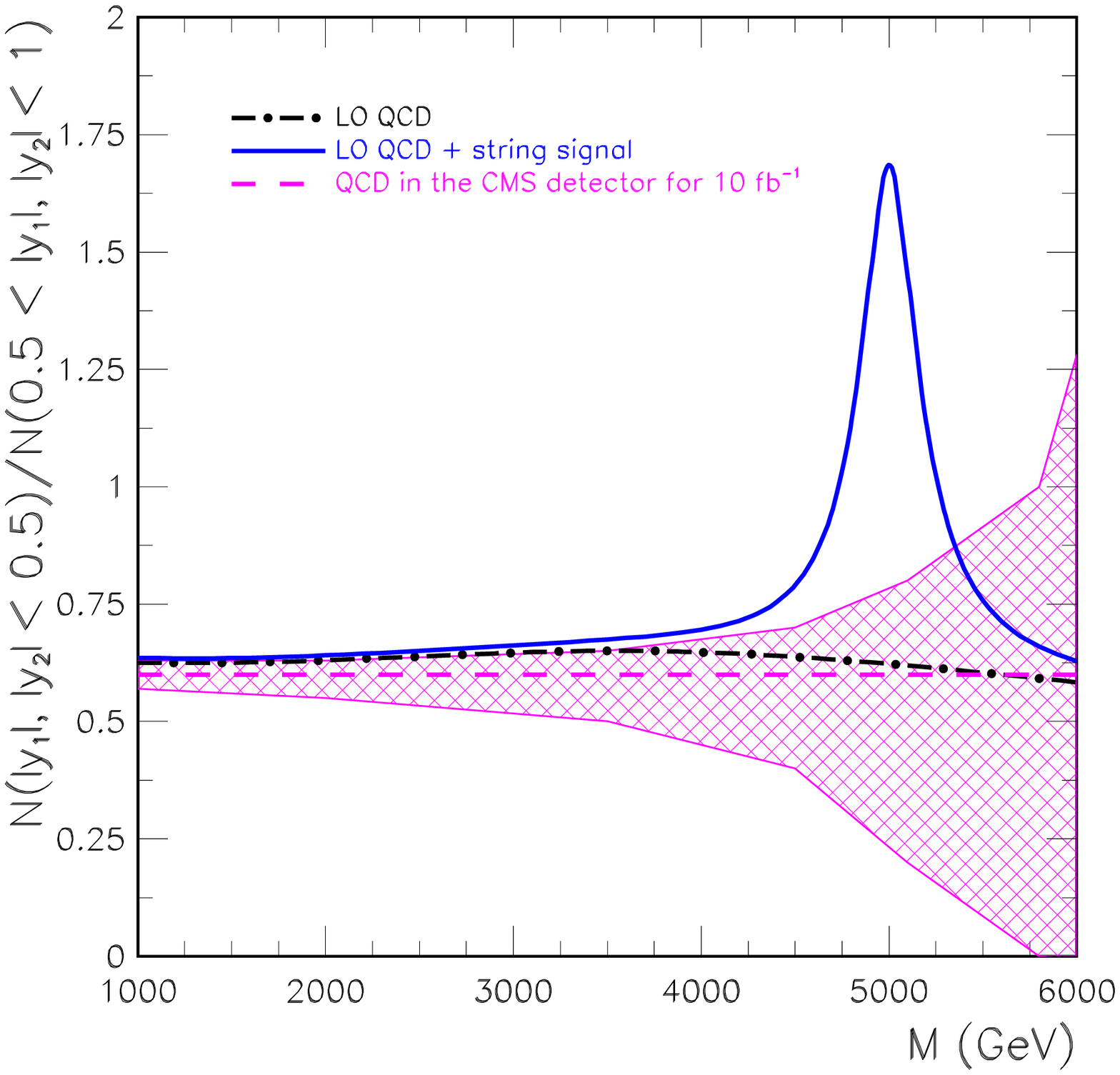}{0.99}
\end{minipage}
\hfill
\begin{minipage}[t]{0.49\textwidth}
\postscript{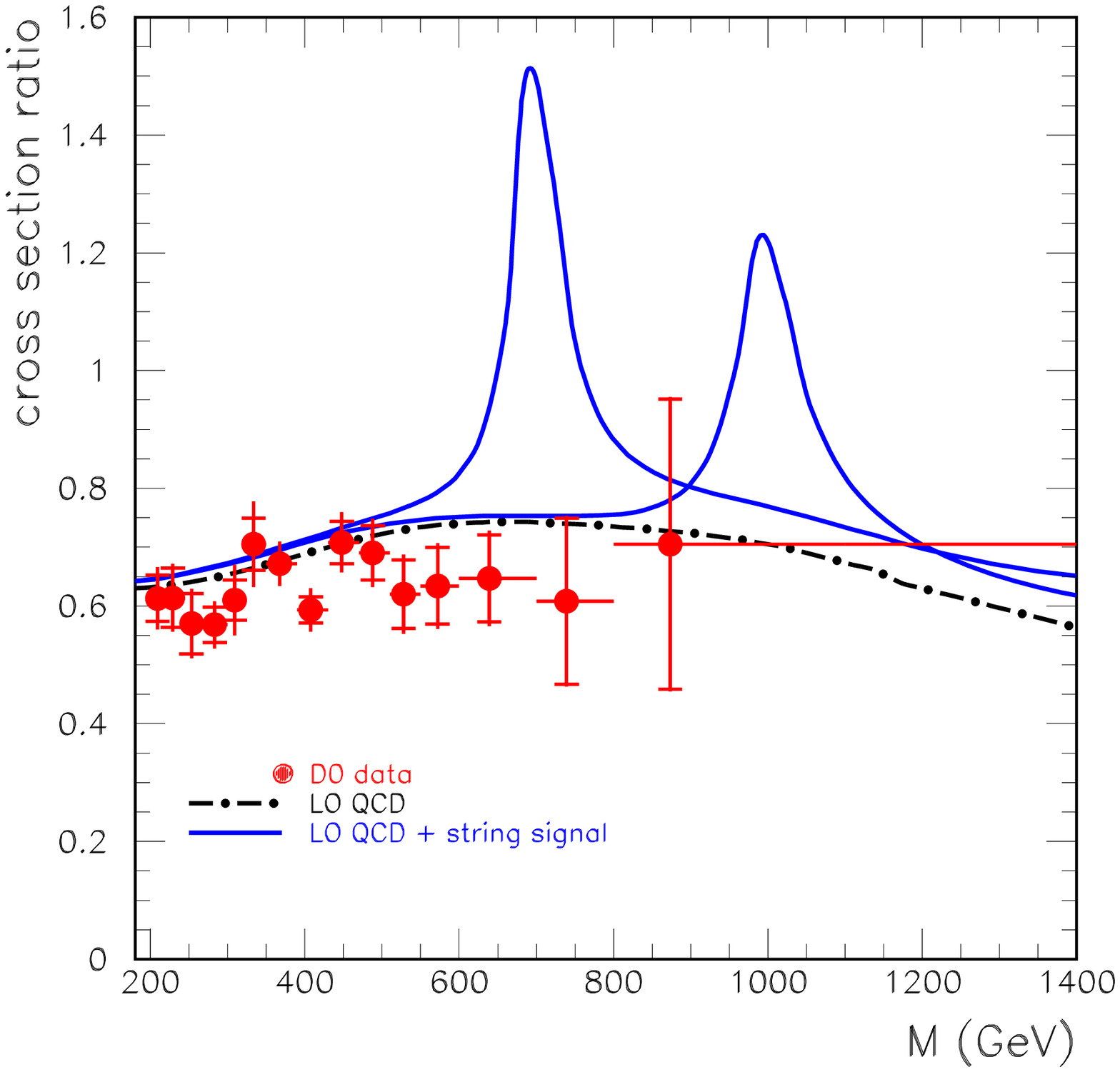}{0.99}
\end{minipage}
\caption{{\em Left panel:} For a luminosity of 10~fb$^{-1}$, the expected
  value (dashed line) and statistical error (shaded region) of the
  dijet ratio of QCD in the CMS detector~\cite{Esen} is compared with
  LO QCD (dot-dashed line) and LO QCD plus lowest massive string
  excitation at a scale $M_s = 5$~TeV. {\em Right panel:} Ratio of dijet
  invariant mass cross sections for rapidities in the interval $0<
  |y_1|,|y_2|< 0.5$ and $0.5 <|y_1|, |y_2|< 1$.  The experimental
  points (solid circles) reprtoted by the D\O\
  Collaboration~\cite{Abbott:1998wh} are compared to a LO QCD
  calculation indicated by a dot-dashed line. The ratio of the string
  + LO QCD invariant mass cross section, in the same rapidity
  intervals, is also shown as a solid line for $M_s = 700$~GeV and
  $M_s = 1$~TeV.  The error bars show the statistical and systematic
  uncertainties added in cuadrature, and the crossbar shows the size
  of the statistical error.}
\label{fig:2}
\end{figure}

We now turn to the analysis of the angular distributions. QCD
parton-parton cross sections are dominated by $t$-channel exchanges
that produce dijet angular distributions which peak at small center of
mass scattering angles. In contrast, non--standard contact
interactions or excitations of resonances result in a more isotropic
distribution. In terms of rapidity variables for standard transverse
momentum cuts, dijets resulting from QCD processes will preferentially
populate the large rapidity region, while the new processes generate
events more uniformly distributed in the entire rapidity region. To
analyze the details of the rapidity space the D\O\
Collaboration~\cite{Abbott:1998wh} introduced a new parameter $R$, the
ratio of the number of events, in a given dijet mass bin, for both
rapidities $|y_1|, |y_2| < 0.5$ and both rapidities $0.5 < |y_1|,
|y_2| < 1.0$~\cite{Meade:2007sz}.  In Fig.~\ref{fig:2} we compare the
results from a full CMS detector simulation of the ratio
$R$~\cite{Esen}, with predictions from LO QCD and model-independent
contributions to the $q^*$, $g^*$ and $C^*$ excitations.  For an
integrated luminosity of 10~fb$^{-1}$ the LO QCD contributions with
$\alpha_{\rm QCD} = 0.1$ (corresponding to running scale $\mu \approx
M_s$) are within statistical fluctuations of the full CMS detector
simulation~\cite{note1}. Since one of the purposes of utilizing NLO
calculations is to fix the choice of the running coupling, we take
this agreement as rationale to omit loops in QCD and in string
theory. It is clear from Fig.~\ref{fig:2} that incorporating NLO
calculation of the background and the signal would not significantly
change the large deviation of the string contribution from the QCD
background. It is noteworthy that although not included in the present
analysis, the signal due to $q q \to q q$, with Regge recurrences and
Kaluza-Klein excitations exchanged in the $t$- and $u$-channels, could
yield a small departure from the QCD value of $R$ outside the resonant
region. In an optimistic scenario, measurements of this modification
could shed light on the D-brane structure of the compact space. It
could also serve to differentiate between a stringy origin for the
resonance as opposed to an isolated structure such as a $Z'$, which
would not modify $R$ outside the resonant region~\cite{note2}.

Before closing, we discuss the impact of Tevatron data on generic
D-brane constructions.  To do so, we compute the cross section ratio, in a
given dijet mass bin, for both rapidities satisfying $|y_1|, |y_2| <
0.5$ and $0.5 < |y_1|, |y_2| < 1.0$, respectively.  We use these results to
calculate the ratio $R$, which is shown in Fig.~\ref{fig:2}. We can
essentially read off from the figure that values of $M_s < 700$~GeV
are excluded at an extremely high CL. Unfortunately, the dijet mass
resolution is not enough to clear up the resonant structure for larger
$M_s$.  For example, at first sight the string prediction for $M_s =
1$~TeV seems to depart from the data in the last dijet mass bin.  To
ascertain any possible deviation, we integrated the cross section over
the last inavriant mass bin ($\sigma (0<|y_1|,|y_2| < 0.5) \simeq
125~{\rm fb}$, $\sigma (0.5 < |y_1|,|y_2| < 1) \simeq 147~{\rm
  fb}$) and estimated an average value of $R = 0.86$ that is in
agreement with D\O\ data~\cite{Abbott:1998wh} at the $1\sigma$
level. One should keep in mind that at the Tevatron the valence
quark contribution $q \bar q \to q \bar q$ is presumably an important
component of the signal. However, in spite of its importance, we are
unable to include this contribution without invoking a
specific compactification scenario~\cite{Cullen:2000ef}.

{\em Conclusions:} In D-brane constructions, the full-fledged string
amplitudes supplying the dominant contributions to dijet cross
sections are completely independent of the details of
compactification. If the string scale is in the TeV range, such
extensions of the standard model can be of immediate phenomenological
interest. In this Letter we have made use of the amplitudes evaluated
near the first resonant pole to determine the discovery potential at
LHC for the first Regge excitations of the quark and gluon. We have
found that, remarkably, the reach of LHC ($S/N = 210/42$)
after a few years of running can be as high as
6.8~TeV~\cite{Antoniadis}. Even after the first 100~pb$^{-1}$ of
integrated luminosity, string scales as high as 4.0~TeV can be
discovered with $S/N = 55/6$. For string scales as high as 5.0
TeV, observations of resonant structures in $pp\rightarrow \gamma~ +$
jet can provide interesting corroboration for stringy physics at the
TeV-scale.

\section*{Acknowledgments}
L.A.A.\ is supported by the U.S. National Science Foundation and the
UWM Research Growth Initiative.  H.G.\ is supported by the
U.S. National Science Foundation Grants No PHY-0244507 and
PHY-0757959.  The research of D.L.\ and St.St.\ is supported in part
by the European Commission under Project MRTN-CT-2004-005104.  The
research of T.R.T.\ is supported by the U.S.  National Science
Foundation Grants PHY-0600304, PHY-0757959 and by the Cluster of
Excellence ``Origin and Structure of the Universe'' in Munich,
Germany.  He is grateful to Arnold Sommerfeld Center for Theoretical
Physics at Ludwig--Maximilians--Universit\"at, and to
Max--Planck--Institut f\"ur Physik in M\"unchen, for their kind
hospitality. Any opinions, findings, and conclusions or
recommendations expressed in this material are those of the authors
and do not necessarily reflect the views of the National Science
Foundation.


\begin{thebibliography}{99}

\bibitem{Anchordoqui:2007da}
  L.~A.~Anchordoqui, H.~Goldberg, S.~Nawata and T.~R.~Taylor,
  Phys.\ Rev.\ Lett.\  {\bf 100}, 171603 (2008)
  [arXiv:0712.0386 [hep-ph]].

\bibitem{Anchordoqui:2008ac}
  L.~A.~Anchordoqui, H.~Goldberg, S.~Nawata and T.~R.~Taylor,
  Phys. Rev. D {\bf 78}, 016005 (2008)
  [arXiv:0804.2013 [hep-ph].]

\bibitem{Anchordoqui:2008hi}
  L.~A.~Anchordoqui, H.~Goldberg and T.~R.~Taylor,
  arXiv:0806.3420 [hep-ph]. In that paper, the particles designated as  $G,\ G^*,\ C^{0*}$
  correspond to $g,\ g^*,\ C^*$ in the present paper.

\bibitem{stringhunter} D.\ L\"ust, S.\ Stieberger and T.R.\ Taylor,
arXiv:0807.3333 [hep-th].


  \bibitem{add}
  N.~Arkani-Hamed, S.~Dimopoulos and G.~R.~Dvali,
  Phys.\ Lett.\  B {\bf 429}, 263 (1998)
  [arXiv:hep-ph/9803315].
\bibitem{ant}
  I.~Antoniadis, N.~Arkani-Hamed, S.~Dimopoulos and G.~R.~Dvali,
  Phys.\ Lett.\  B {\bf 436}, 257 (1998)
  [arXiv:hep-ph/9804398].

\bibitem{inter} For a review, see:
  R.~Blumenhagen, B.~Kors, D.~L\"ust and S.~Stieberger,
  Phys.\ Rept.\  {\bf 445}, 1 (2007)
  [arXiv:hep-th/0610327].


\bibitem{gabriele}
G. Veneziano, Nouvo Cimento A {\bf 57}, 190 (1968).


\bibitem{CMS}
  A.~Bhatti {\it et al.},
  arXiv:0807.4961 [hep-ex].

\bibitem{Pumplin:2002vw}
  J.~Pumplin, D.~R.~Stump, J.~Huston, H.~L.~Lai, P.~Nadolsky and W.~K.~Tung,
  JHEP {\bf 0207}, 012 (2002)
  [arXiv:hep-ph/0201195].


\bibitem{Ball:2007zza}
  G.~L.~Bayatian {\it et al.}  [CMS Collaboration],
  J.\ Phys.\ G {\bf 34} 995 (2007);
  W.~W.~Armstrong {\it et al.}  [ATLAS Collaboration],
  CERN/LHCC 94-43.



\bibitem{pizero} The approximate equality of the background due to
  misidentified $\pi^0$'s and the QCD background, across a range of
  large $p_T^\gamma$ as implemented in Ref.~\cite{Anchordoqui:2008ac},
  is maintained as an approximate equality over a range of invariant
  $\gamma$-jet invariant masses with the rapidity cuts imposed.


\bibitem{Abbott:1998wh}
  B.~Abbott {\it et al.}  [D0 Collaboration],
  Phys.\ Rev.\ Lett.\  {\bf 82}, 2457 (1999)
  [arXiv:hep-ex/9807014].



\bibitem{Meade:2007sz} An illustration of the use of this parameter in
  a heuristic model where standard model amplitudes are modified by a
  Veneziano formfactor has been presented by P.~Meade and L.~Randall,
  JHEP {\bf 0805}, 003 (2008)
  [arXiv:0708.3017 [hep-ph]].


\bibitem{Esen} The synthetic population was generated with Pythia,
  passed through the full CMS detector simulation and reconstructed
  with the ORCA reconstruction package;  
  S. Esen and R. Harris,
  CMS Note 2006/071.

\bibitem{note1} Note that the string scale is an optimal choice of the
  running scale which should normally minimize the role of higher loop
  corrections.

\bibitem{note2} Because of the high multiplicity of the angular
  momenta (up to $J=2$), the rapidity distribution of the decay
  products of string excitations would differ
  significantly from those following decay of a $Z'$ with $J=1$.  With
  higher statistics, isolation of lowest massive Regge excitations from 
  Kaluza-Klein replicas (with $J=2$) may also be possible.

\bibitem{Cullen:2000ef}  Four fermion channels were used
 previously to obtain model dependent bounds on $M_s$. See e.g.,
  S.~Cullen M.~Perelstein and M.~E.~Peskin,
  Phys.\ Rev.\  D {\bf 62}, 055012 (2000)
  [arXiv:hep-ph/0001166];
  P.~Burikham, T.~Figy and T.~Han,
  Phys.\ Rev.\  D {\bf 71}, 016005 (2005)
  [Erratum-ibid.\  D {\bf 71}, 019905 (2005)]
  [arXiv:hep-ph/0411094];
  K.~Cheung and Y.~F.~Liu,
  Phys.\ Rev.\  D {\bf 72}, 015010 (2005)
  [arXiv:hep-ph/0505241].

\bibitem{Antoniadis} This intersects with the range of string scales
consistent with correct weak mixing angle found in the model of
  I.~Antoniadis, E.~Kiritsis and T.~N.~Tomaras,
  Phys.\ Lett.\  B {\bf 486}, 186 (2000)
  [arXiv:hep-ph/0004214].




\end{thebibliography}
\end{document}